\documentclass[aps,
pre,
notitlepage,
]{revtex4-1}
\usepackage{dcolumn}
\usepackage{bm}%

\usepackage{setspace}
\usepackage{fullpage}
\usepackage{dsfont}
\usepackage[basic,physics]{circ}

\usepackage{amssymb}
\usepackage{amsmath}
\usepackage{amsfonts}
\usepackage{eulervm}
\usepackage{graphicx,subfigure}
\usepackage{epstopdf}
\usepackage{hyperref}
\usepackage{color}
\usepackage{eso-pic}

\newcommand{\be}{\begin{equation}}
\newcommand{\ee}{\end{equation}}
\newcommand\numberthis{\addtocounter{equation}{1}\tag{\theequation}}

\newcommand{\infectious}{\text{\small I}}
\newcommand{\susceptible}{\text{\small S}}
\newcommand{\recovered}{\text{\small R}}
\newcommand{\exposed}{\text{\small E}}

\newcommand{\rDMP}{\ensuremath{\mathrm{rDMP}}}

\newcommand{\pair}{\mathrm{pair}}
\newcommand{\wane}{\rho}
\newcommand{\itor}{\rho}
\newcommand{\rtos}{\gamma}
\newcommand{\etoi}{\varepsilon}
\newcommand{\J}{\mathbf{J}}
\renewcommand{\A}{\mathbf{A}}
\newcommand{\B}{\mathbf{B}}
\newcommand{\id}{\mathds{1}}

\begin{document}
\title{A message-passing approach for recurrent-state epidemic models on networks}
\date{\today}
\author{Munik Shrestha}
\affiliation{University of New Mexico, Albuquerque, NM 87131, USA}
\affiliation{Santa Fe Institute, 1399 Hyde Park road, Santa Fe, NM 87501, USA}
\author{Samuel V. Scarpino}
\affiliation{Santa Fe Institute, 1399 Hyde Park road, Santa Fe, NM 87501, USA}
\author{Cristopher Moore}
\affiliation{Santa Fe Institute, 1399 Hyde Park road, Santa Fe, NM 87501, USA}

\begin{abstract}
Epidemic processes are common out-of-equilibrium phenomena of broad interdisciplinary interest. Recently, dynamic message-passing (DMP) has been proposed as an efficient algorithm for simulating epidemic models on networks \cite{KarrerNewman_2010,miller_2011, ShresthaMoore_2014_1, Lokhov_2014, Altarelli_2013_L}, and in particular for estimating the probability that a given node will become infectious at a particular time. To date, DMP has been applied exclusively to models with one-way state changes, as opposed to models like SIS (susceptible-infectious-susceptible) and SIRS (susceptible-infectious-recovered-susceptible) where nodes can return to previously inhabited states. Because many real-world epidemics can exhibit such recurrent dynamics, we propose a DMP algorithm for complex, recurrent epidemic models on networks.  Our approach takes correlations between neighboring nodes into account while preventing causal signals from backtracking to their immediate source, and thus avoids ``echo chamber effects" where a pair of adjacent nodes each amplify the probability that the other is infectious.  We demonstrate that this approach well approximates results obtained from Monte Carlo simulation and that its accuracy is often superior to the pair approximation (which also takes second-order correlations into account).  Moreover, our approach is more computationally efficient than the pair approximation, especially for complex epidemic models: the number of variables in our DMP approach grows as $2mk$ where $m$ is the number of edges and $k$ is the number of states, as opposed to $mk^2$ for the pair approximation. We suspect that the resulting reduction in computational effort, as well as the conceptual simplicity of DMP, will make it a useful tool in epidemic modeling, especially for inference tasks where there is a large parameter space to explore.
\end{abstract}
\maketitle

\section{Introduction}
\label{sec:intro}

Mathematical models of epidemic processes are intrinsically non-linear and multiplicative.  These models include the spread of disease~\cite{Bailey75,AndersonMay91}, transmission of social behaviors \cite{Gran78,Gran73,JMiller04,Goncalves}, cascades of banking failures \cite{BMay2011,CSMF2012}, forest fires \cite{BakChen1,Drossel1,Grassberger1}, the propagation of marginal probabilities in constraint satisfaction problems~\cite{MezardMontanari,MooreMertens} and the dynamics of magnetic and glassy systems~\cite{RMorris1}.

The classical approach to modeling epidemics, such as the SIR model where each node is Susceptible, Infectious, or Recovered, assumes that at any given time each individual exists in a single state or ``compartment''~\cite{Bailey75, AndersonMay91}. To make these models analytically tractable, it is often assumed that the population is well mixed, so that interaction between any two individuals is equally likely; in physical terms, we assume the model is mean-field (also known as mass-action mixing in the epidemiology literature).  
Despite this unrealistic assumption, mean-field models capture some essential features of epidemics, such as a threshold above which we have an endemic phase with a non-zero fraction of infected individuals, and below which we have outbreaks of size $o(n)$ so that the equilibrium fraction of infected individuals is zero.

In reality, contacts between individuals in the population are often highly structured, with some pairs of individuals much more likely to interact than others due to location or demographics~\cite{Dunbar, Goncalves}. To relax the mean-field assumption, while retaining some measure of tractability, we can assume that individuals interact on a network, whose structure captures the heterogeneity in the population~\cite{Meyers2007, Newman_Network_Book}. However, replacing the mean-field approximation with a contact network substantially increases a model's complexity.

One reasonable goal is to compute the one-point marginals, e.g., for each node $i$ the probability $\infectious_i(t)$ that $i$ is infectious at time $t$.  In addition to being of direct interest, these marginals help us perform tasks such as inferring the originator of an epidemic, determining an optimal set of nodes to immunize in order to minimize the final size of an outbreak, or calculating the probability that an entire group of nodes will remain uninfected after a fixed time \cite{Lokhov_2013, Altarelli_2014_1, Altarelli_2013_1, Altarelli_2013_2, Altarelli_2011_1}.  

We can always compute these marginals by performing Monte Carlo experiments.  However, since we need to perform many independent trials in order to collect good statistics, this is computationally expensive on large networks.  This problem is compounded if we need to scan through parameter space, or if we want to explore many different initial conditions, vaccination strategies, etc.  Therefore, it would be desirable to compute these marginals using, say, a system of differential equations, with variables that directly model the probabilities of various events.

The most naive way to do this, as we review below, uses the one-point marginals themselves as variables.  However, this approach completely ignores correlations between nodes.  At the other extreme, to model the system exactly, we would need to keep track of the entire joint distribution: but if there are $n$ individuals, each of which can be in one of $k$ states, this results in a coupled system with $k^n$ variables.  This exponential scaling quickly renders most models computationally intractable, even on moderately sized networks.  

In between these two extremes, we can approximate the joint distribution by ``moment closure,'' assuming that higher-order marginals can be written in terms of lower-order ones.  This gives a hierarchy of increasingly accurate (and computationally expensive) approximations, familiar in physics as cluster expansions.  At the first level of this hierarchy we assume that the nodes are uncorrelated, and approximate two-point marginals such as $[\infectious_i(t) \wedge \infectious_j(t)]$ (the probability that $i$ and $j$ are both infectious at time $t$) as $\infectious_j(t) \infectious_j(t)$.  At the second level, commonly referred to in the epidemiology literature as the pair approximation, we close the hierarchy at the level of pairs $[ \infectious_i(t) \wedge \infectious_j(t) ]$ by assuming that three-point correlations can be factored in terms of two-point correlations.  For a comprehensive review of these methods, see~\cite{Newman_Network_Book,Mason-review-2015}.

In this paper, we study an alternative method, namely \emph{Dynamic Message-Passing} (DMP).  As in belief propagation \cite{JPearl1,Decelle2011}, here variables or ``messages'' are defined on a network's directed edges: for instance, $\infectious_{j \to i}$ denotes the probability that $j$ was infected by one of its neighbors other than $i$, so that the epidemic might spread from $j$ to $i$. However, unlike belief propagation, where the posterior distributions are updated according to Bayes' rule, here we write differential equations for the messages over time. 

For many epidemic models, such as SI (susceptible-infectious), SIR (susceptible-infectious-recovered) and SEIR (susceptible-exposed-infectious-recovered), only one-way state changes can occur. For example, in the SIR model, once an individual has left the Susceptible class and become Infectious, they cannot return to being Susceptible; once they become Recovered, they are immune to future infections, and might as well be Removed.  For these non-recurrent models, DMP is known to be be an efficient algorithm to estimate $\infectious_i(t)$, and it is exact on trees~\cite{KarrerNewman_2010}; it can also be applied to threshold models~\cite{ShresthaMoore_2014_1,Altarelli_2013_L,Lokhov_2014} and used for inference~\cite{Lokhov_2013}. 

However, for many real-world diseases individuals can return to previously inhabited states.  In these \emph{recurrent} models, such as SIS (susceptible-infectious-susceptible), SIRS (susceptible-infectious-recovered-susceptible), and SEIS (susceptible-exposed-infectious-susceptible), individuals can cycle through the states multiple times, giving multiple waves of infection traveling through the population. The most obvious examples of recurrent models are seasonal influenza, where due to the evolution of the virus individuals are repeatedly infected during their lifetime~\cite{earn2002ecology}, vaccination where protective immunity wanes over time~\cite{gomes2004infection}, and diseases curable by treatment which does not result in antibody-mediated immunity, such as gonorrhea~\cite{golden2005effect}. In all three cases, individuals leave the Susceptible class, only to return at some point in the future (although for influenza, it is worth mentioning that if the evolutionary rate of the virus is functionally related to the number of susceptible individuals, then the recovery rate may not be independent from the state of one's neighbors.)  Unfortunately, the DMP approach of~\cite{KarrerNewman_2010} cannot be directly extended to recurrent models, since their equations for messages only track the first time an individual makes the transition to a given state.

The purpose of this paper is to develop a novel DMP algorithm for recurrent models of epidemics on networks, which we call \rDMP.  We will show that \rDMP\ gives very good approximations for marginal probabilities on networks, and is often more accurate than the pair approximation. Moreover, whereas the pair approximation requires keeping track of $mk^2$ variables, if there are $m$ edges and $k$ states per node, \rDMP\ requires just $2mk$ variables.  For complex models where $k$ is large---for instance, for diseases with multiple stages of infection or immunity, or multiple-disease epidemics where one disease makes individuals more susceptible to another one---this gives a substantial reduction in the computational effort required.  Finally, the \rDMP\ approach is conceptually simple, making it easy to write down the system of differential equations for a wide variety of epidemic models.

\section{Message-Passing and Preventing the Echo Chamber Effect}
\label{sec:echo}

 \begin{figure}
\centering
\includegraphics[width=1.5in]{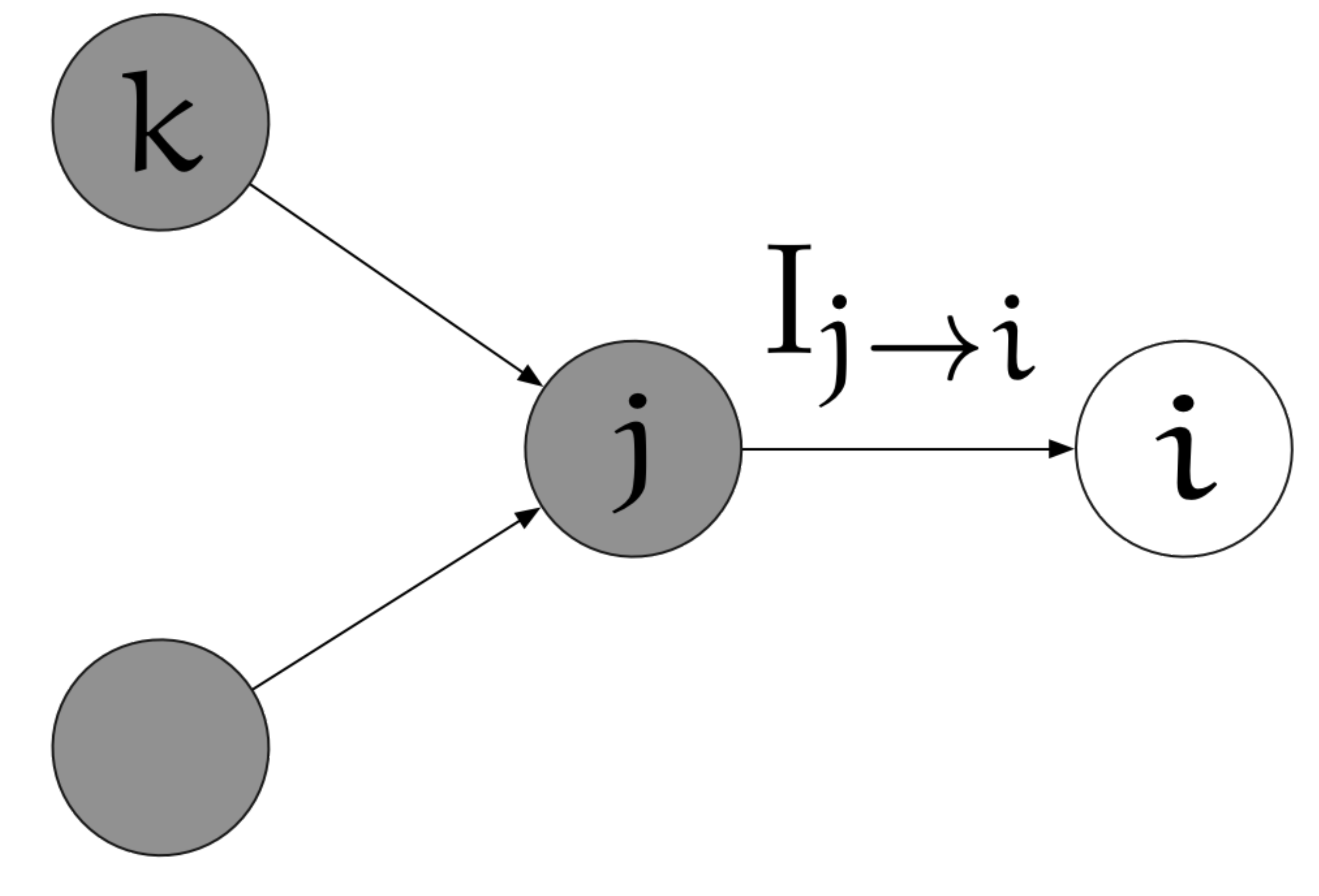}
\caption{We define messages on the directed edges of a network to carry causal information of the flow of contagion, e.g. $\infectious_{j \to i}$ is the probability that $j$ is Infectious because it received the infection from a neighbor $k$ other than $i$. This prevents effects from immediately backtracking to the node they came from, and avoids ``echo chamber'' infections.}
\label{fig:messages}
\end{figure}

\begin{figure}
\centering
\includegraphics[width=3.1in]{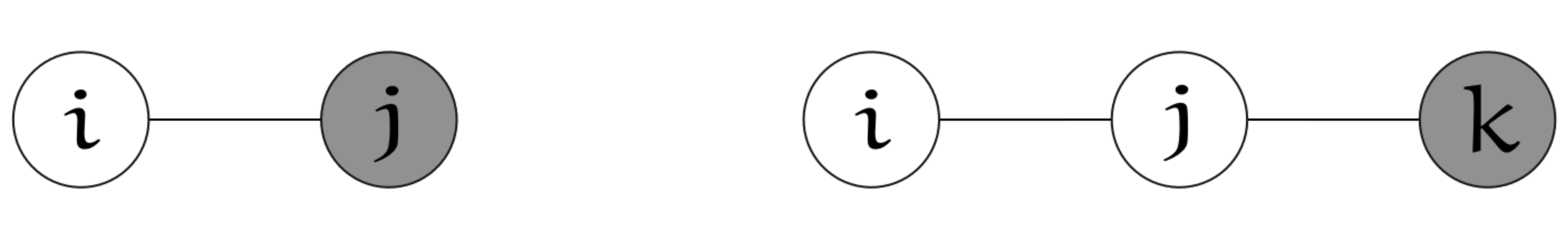}
\caption{Two simple, yet illustrative, cases of networks, where the darker node is initially Infectious. As we discuss, in these simple cases one can see the motivation for our approach to prevent infection signals from backtracking to the node it immediately came from.}
\label{fig:echo}
\end{figure}

As shown in Fig.~\ref{fig:messages}, the variables of \rDMP\ are messages along directed edges of the network (in addition to one-point marginals).  For instance, $\infectious_{j \to i}$ is the probability that $j$ is Infectious because it was infected by one of its other neighbors $k$.  The intuition behind this is the following, where we take the SIS model as an example.  If $i$ is Susceptible, the rate at which $j$ will infect $i$ is proportional to the probability $\infectious_j$ that $j$ is infected.  But when computing this rate, we only include the contribution to $\infectious_j$ that comes from neighbors other than $i$.  In other words, we deliberately neglect the event that $j$ receives the infection from $i$, and immediately passes it back to $i$, even if $i$ has become Susceptible in the intervening time.

This choice avoids a kind of ``echo chamber'' effect, where neighboring nodes artificially amplify each others' probability of being Infectious.  For instance,  consider a simple but pathological case of the SI model where there are only two nodes in the graph, $i$ and $j$, with an edge between them as shown in Fig.~\ref{fig:echo}.  If the transmission rate is $\lambda$, and if we assume the nodes are independent (i.e., if we use first-order moment closure) we obtain the following differential equations, 
\begin{align*}
\frac{d\infectious_i}{dt} &= \lambda \susceptible_i \infectious_j \\
\frac{d\infectious_j}{dt} &= \lambda \susceptible_j \infectious_i \, , \numberthis \label{eq:node_twonode}
\end{align*}
where $\susceptible_i(t) = 1-\infectious_i(t)$ and similarly for $j$.  

Now suppose that $j$ is initially Infectious with probability $\delta$, and that $i$ is initially Susceptible, i.e., $\infectious_j(0)=\delta$ and $\infectious_i(0)=0$.  Since in the SI model nodes never recover, the infection will eventually spread from $j$ to $i$, but only if $i$ was Infectious in the first place.  Thus the marginals $\infectious_i(t)$ and $\infectious_j(t)$ should tend to $\delta$ as $t \to \infty$.

However, integrating Eq.~\eqref{eq:node_twonode} gives a different result.  Once $\infectious_i$ becomes positive, $d\infectious_j/dt$ becomes positive as well, allowing $i$ to infect $j$ with the infection that it received from $j$ in the first place.  As a result, $\infectious_j(t)$ approaches $1$ as $t \to \infty$.  Thus the ``echo chamber'' between $i$ and $j$ leads to the absurd result that $j$ eventually becomes Infectious, even though with probability $1-\delta$ there was no initial infection in the system.

In the \rDMP\ approach, we fix this problem by replacing $\infectious_i$ and $\infectious_j$ with the messages they send each other,
\begin{align*}
\frac{d\infectious_i}{dt} &= \lambda \susceptible_i \infectious_{j \to i} \, , \\
\frac{d\infectious_j}{dt} &= \lambda \susceptible_j \infectious_{i \to j} \, ,
\end{align*}
so that $i$ can only infect $j$ if $i$ received the infection from some node other than $j$.  (Below we give the equations on a general network, including the time derivatives of the messages.)  In this example, there are no other nodes, so if $\infectious_{j \to i}(0) = \delta$ and $\infectious_{i \to j}(0) = 0$, then $\infectious_j(t) = \delta$ for all $t$ as it should be.  

Note that we do not claim that \rDMP\ is exact in this case.  In particular, as in~\eqref{eq:node_twonode}, $\infectious_i(t)$ tends to $1$ as $t \to \infty$.  This is because, unlike the system of~\cite{KarrerNewman_2010}, \rDMP\ assumes that the events that $j$ infects $i$ at different times are independent.

In this two-node example, of course, the pair approximation is exact, since it maintains separate variables such as $[\susceptible_j \wedge \infectious_k]$ for each of the joint states of the two nodes.  However, the pair approximation is subject to other forms of the echo chamber effect.  Consider a network with three nodes, as in Fig.~\ref{fig:echo} (right), where $j$ is a common neighbor of $i$ and $k$.  The pair approximation assumes that, conditioned on the state of $j$, the states of $i$ and $k$ are independent; however, in a recurrent epidemic model, $i$ and $k$ could be correlated, for instance if $j$ infected them both and then returned to the Susceptible state.  As a result, the pair approximation is vulnerable to a distance-two echo chamber, where $i$ and $k$ infect each other through $j$.  As in the two-node case, \rDMP\ prevents this.

Preventing backtracking completely may seem like a strong assumption, and in recurrent models it is \emph{a priori} possible, for instance, for a node to re-infect the neighbor it was infected by.  Despite the well-documented importance of recurrent infections for diseases including (but certainly not limited to) seasonal influenza~\cite{earn2002ecology}, Plasmodium malaria~\cite{jeffery1966epidemiological}, and urinary tract infections~\cite{conway2007recurrent}, little is known about the source of recurrent infections.  For certain sexually transmitted diseases such as gonorrhea~\cite{golden2005effect} and repeated ringworm infections~\cite{drusin2000nosocomial}, there is evidence that backtracking plays a significant role; on the other hand, it may be that recurrent infections are caused by different strains, each of which is acting essentially without backtracking.  Thus while our non-backtracking assumption is clearly invalid in some cases, we believe it is a reasonable approach for most recurrent state infections.

\section{The \rDMP\ Equations for the SIS, SIRS, and SEIS Models}
\label{sec:DMP}

In this section, we illustrate the \rDMP\ approach for several recurrent epidemic models.  We start with the simplest one: in the SIS model, each node is either Infectious ($\infectious$) or Susceptible ($\susceptible$). Infectious nodes infect their Susceptible neighbors at rate $\lambda$, and their infections wane back into the Susceptible state at rate $\wane$.  We denote the probability that that node $i$ is Infectious or Susceptible by $\infectious_i$ and $\susceptible_i$ respectively. The objective then is to efficiently and accurately compute these probabilities as a function of time $t$. 

We define variables or ``messages'' that live on the directed edges $(i,j)$ of the network.  The directed nature of these messages prevent infection from backtracking from an Infectious node back to its infection source, e.g., if node $i$ infects node $j$, then we prevent $j$ from re-infecting $i$.  In addition to tracking the one-point marginal $\infectious_j$, we define a message $\infectious_{j \rightarrow i}$ from $j$ to $i$ as the probability that $j$ is in the Infectious state as a result of being infected from one of its neighbors other than $i$.  Given these incoming messages, the rate at which $\infectious_i$ evolves in time is given by
\be
\label{dmp_node_sis}
\frac{d\infectious_i }{dt} = -\wane \infectious_i  + \lambda  \susceptible_i  \sum_{j \in \partial i} \infectious_{j \to i}, 
\ee
where $\partial i$ denotes the neighbors of $i$.  
Similarly, the rate at which $\infectious_{j \rightarrow i}$ evolves in time is given by 
\begin{align}\label{dmp_edge_sis}
& \frac{d\infectious_{j \to i}}{dt}= -\wane \infectious_{j \to i} +\lambda \susceptible_{j}  \sum_{k \in \partial j  \setminus i}\infectious_{k \to j},
\end{align}
where $k \in \partial j  \setminus i$ denotes the neighbors of $j$ excluding $i$.  

For the SIRS model, we let $\itor$ and $\rtos$ denote the transition rates from Infectious to Recovered and from Recovered to Susceptible respectively.  Then the \rDMP\ system for the SIRS model is given by 
\begin{align}
&\frac{d\infectious_{j \to i}}{dt}= -\itor \infectious_{j \to i} +\lambda \susceptible_{j}  \sum_{k \in \partial j  \setminus i}\infectious_{k \to j},
\end{align}
which is coupled with the one-point marginals through
\begin{align*}
\frac{d\susceptible_i }{dt} &= \rtos \recovered_i - \lambda \susceptible_i  \sum_{j \in \partial i} \infectious_{j \to i} \\
\frac{d\infectious_i }{dt} &= -\itor \infectious_i  + \lambda  \susceptible_i  \sum_{j \in \partial i} \infectious_{j \to i} \\
\frac{d\recovered_i }{dt} &= \itor \infectious_i - \rtos \recovered_i \, . 
\numberthis \label{dmp_node_sirs}
\end{align*}

In the SEIS model, upon becoming exposed to an infected neighbor, Susceptible nodes first go through a latent period called the Exposed state.  In this state, individuals are infected but not yet Infectious.  Exposed nodes become Infectious at the rate $\etoi$, and Infectious nodes again wane back to Susceptible at rate $\wane$.  The \rDMP\ system for the SEIS model is
\begin{align*}
\frac{d\exposed_{j \to i}}{dt} 
&= -\etoi \exposed_{j \to i} + \lambda \susceptible_j  \sum_{k \in \partial j \setminus i} \infectious_{k \to j}, \\
\frac{d\infectious_{j \to i}}{dt} 
&= -\wane \infectious_{j \to i} + \etoi \exposed_{j \to i} \, , 
\numberthis \label{dmp_edge_sirs}
\end{align*}
which is coupled with the one-point marginals as
\begin{align*}
\frac{d\susceptible_i }{dt} 
&= \wane \infectious_i-\lambda  \susceptible_i  \sum_{j \in \partial i} \infectious_{j \to i} \\
\frac{d\infectious_i }{dt} 
&=-\wane\infectious_i  + \etoi\exposed_i \\
\frac{d\exposed_i }{dt}
&= -\etoi \infectious_i + \lambda \susceptible_i \sum_{j \in \partial i} \infectious_{j \to i} \, .
\numberthis\label{dmp_node_seis}
\end{align*}
Note that here we track messages for the Exposed state, in addition to one-point marginals, since they act as precursors for the Infectious messages.  There is no need to track messages for the Susceptible state, since it does not cause state changes in its neighbors.  

Generalizing these equations to more complex epidemic models with $k$ different states, as opposed to three or four, is straightforward.  Even in a model where every state can cause state changes in its neighbors---for instance, where having Susceptible neighbors speeds up the rate of recovery, or where Exposed nodes can also infect their neighbors at a lower rate---the total number of variables we need to track in a network with $n$ nodes and $m$ edges is at most $2mk$ in addition to the $nk$ one-point marginals.  In contrast, the pair approximation requires $mk^2$ states to keep track of the joint distribution of every neighboring pair. 

\section{Experiments in Real and Synthetic Networks}
\label{sec:experiments}

In this section we report on numerical experiments for \rDMP\ for the SIS and SIRS models on real and synthetic networks.  As a performance metric, we use the average $L_1$ error per node between the marginals computed from \rDMP\ and the true probabilities computed (up to sampling error) using continuous-time Monte Carlo simulations.  That is, 
\be
\label{eq:L1norm}
L_1^{\rDMP} (t) = \frac{1}{n}\sum_i \left| \infectious_i^{\mathrm{MC}} (t)-\infectious_i^{\rDMP}(t) \right| \, ,
\ee
We use this metric to compare the performance of \rDMP\ with the independent-node approximation and the pair approximation, or equivalently first- and second-order moment closure~\cite{Newman_Network_Book,Mason-review-2015}.  As we will see, for a wide range of parameters, \rDMP\ is more accurate than either of these approaches, even though it is computationally easier than the pair approximation.

\begin{figure*}
\centering
\mbox{\subfigure{\includegraphics[width=3.2in]{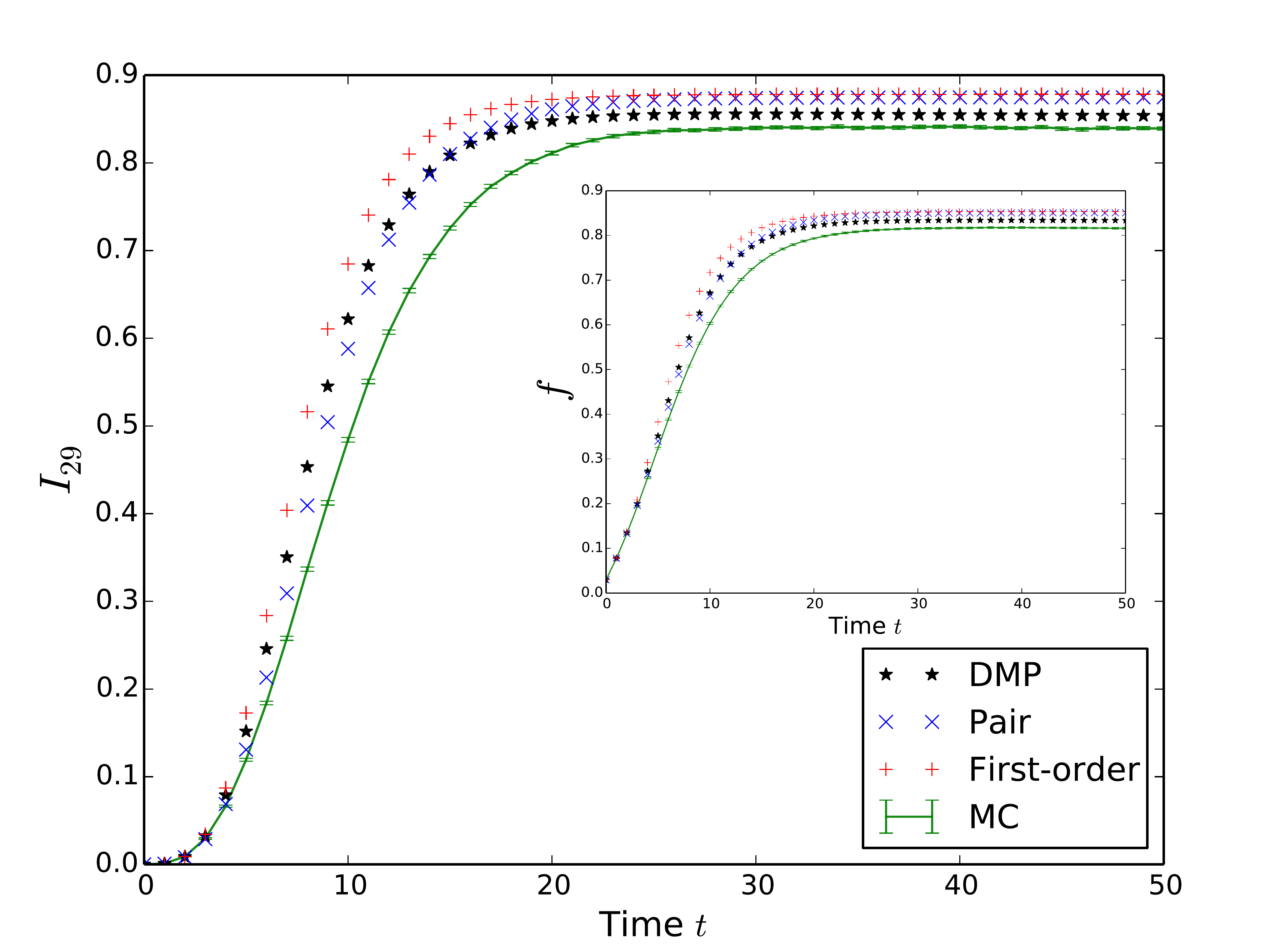}}\quad
\subfigure{\includegraphics[width=3.2in]{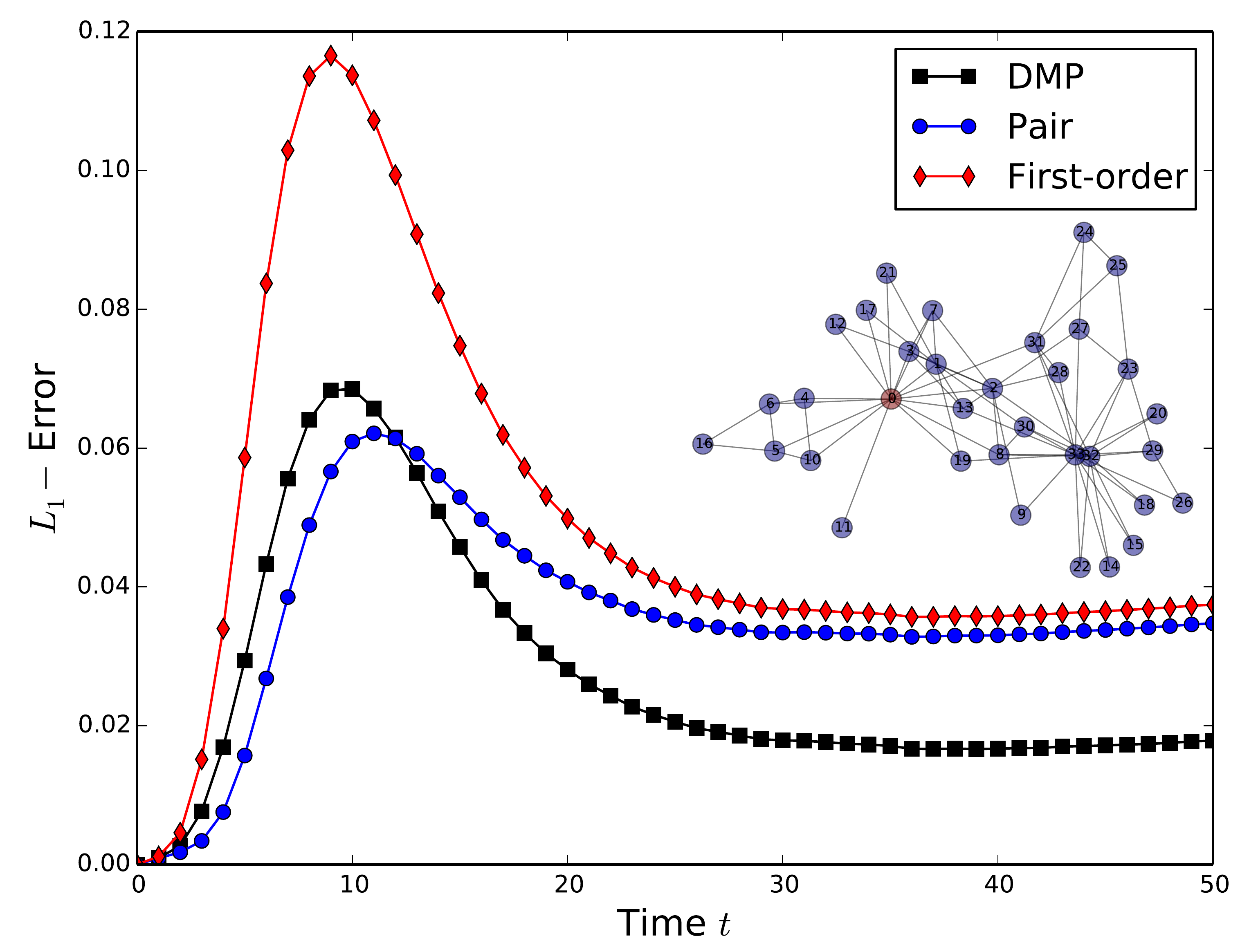}}}
\caption{Results on the SIS model.  On the left, the marginal probability that node 29 in Zachary's Karate club (see inset on right) is Infectious as a function of time.  We compare the true marginal derived by $10^5$ independent Monte Carlo simulations with that estimated by \rDMP, the independent node approximation, and the pair approximation.  On the right is the $L_1$ error, averaged over all nodes; we see that \rDMP\ is the most accurate of the three methods.  Here the transmission rate is $\lambda = 0.1$, the waning rate is $\wane = 0.05$, and vertex $0$ (colored red) was initially infected.}
\label{fig:L1distance}
\end{figure*}

\begin{figure}
\centering
\includegraphics[width=3.2in]{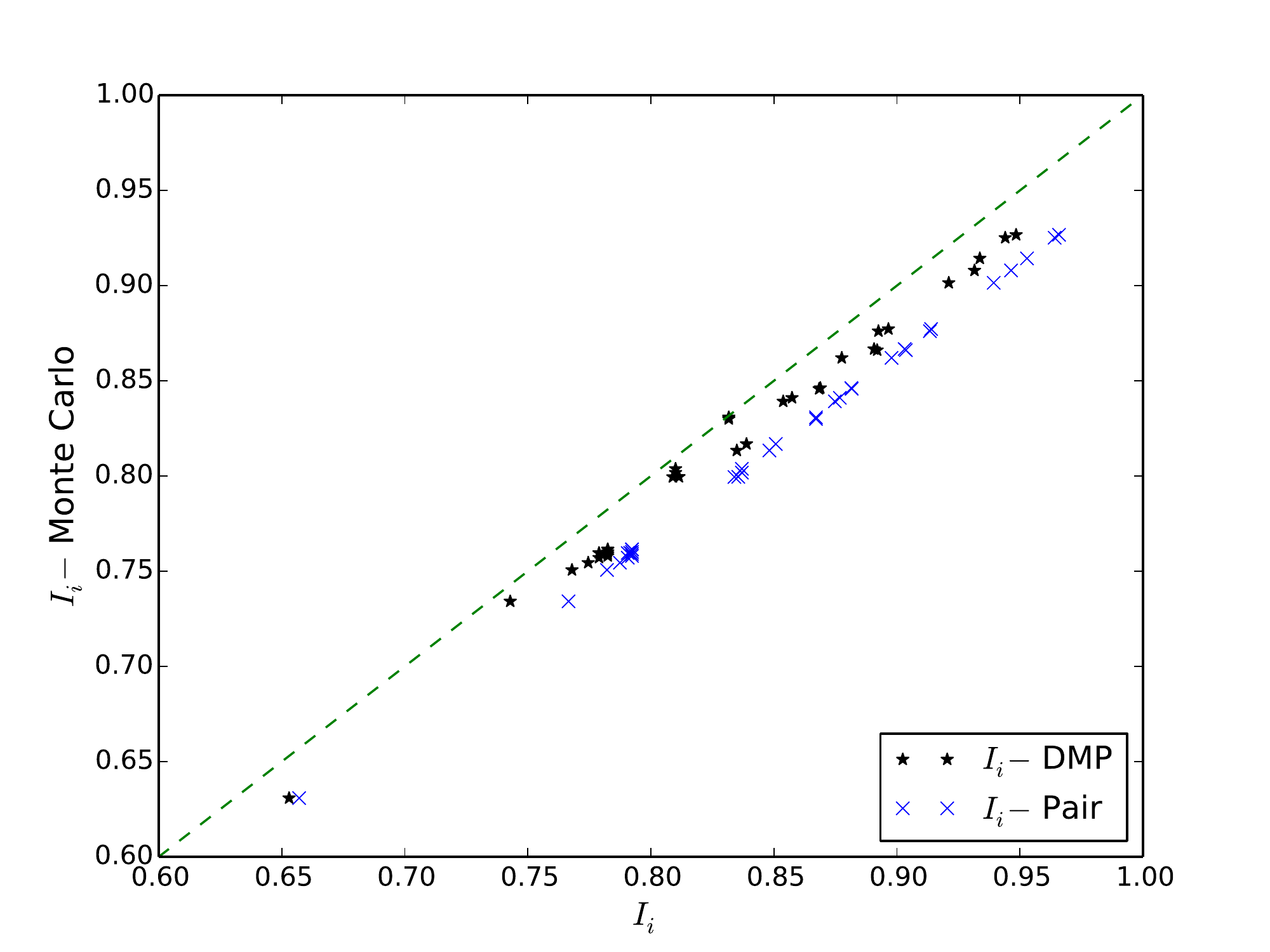}
\caption{A scatterplot of the steady-state marginals $\infectious_i$ for the $n=33$ nodes in Zachary's Karate Club, with the same parameters as in Fig.~\ref{fig:L1distance}.  The vertical axis is the true marginal computed by Monte Carlo simulations; the horizontal axis is the estimated marginals from \rDMP\ (black $\star$) and the pair approximation (blue $\times$).  Both methods overestimate the marginal, but \rDMP\ is closer to the true value (the line $y=x$) for every node.}
\label{fig:scatter-plot}
\end{figure}

In Fig.~\ref{fig:L1distance}, we show results for the SIS model on Zachary's Karate Club~\cite{Zachary}.  On the left, we show the marginal probability that a particular node is Infectious as a function of time, estimated by \rDMP\ and by first- and second-order moment closure, and compared with the true marginals given by Monte Carlo simulation.  On the right, we show the average $L_1$ error for the three methods.  Here $\lambda = 0.1$, $\wane = 0.05$, and the initial condition consists of a single infected node (shown in red in the inset).  The Monte Carlo results were averaged over $10^5$ runs.  We see that \rDMP\ is significantly more accurate than the other two, except at some early times when the pair approximation marginally outperforms \rDMP. 

As a further illustration, in Fig.~\ref{fig:scatter-plot} we show the steady-state marginal $\infectious_i$ for each node $i$ (measured by running the system until $t=50$, at which point $\infectious_i(t)$ is nearly constant), with the same parameters and initial condition as in Fig.~\ref{fig:L1distance}.  We show the true marginal of each node on the $y$-axis, and the marginals estimated by \rDMP\ and the pair approximation on the $x$-axis.  If the estimated marginals were perfectly accurate, the points would fall on the line $y=x$.  Both methods overestimate the marginals to some extent, but \rDMP\ is more accurate than the pair approximation on every node.  Thus \rDMP\ makes accurate estimates of the marginals on individual nodes, as opposed to just the average across the population.

\begin{figure}
\centering
\includegraphics[width=3.2in]{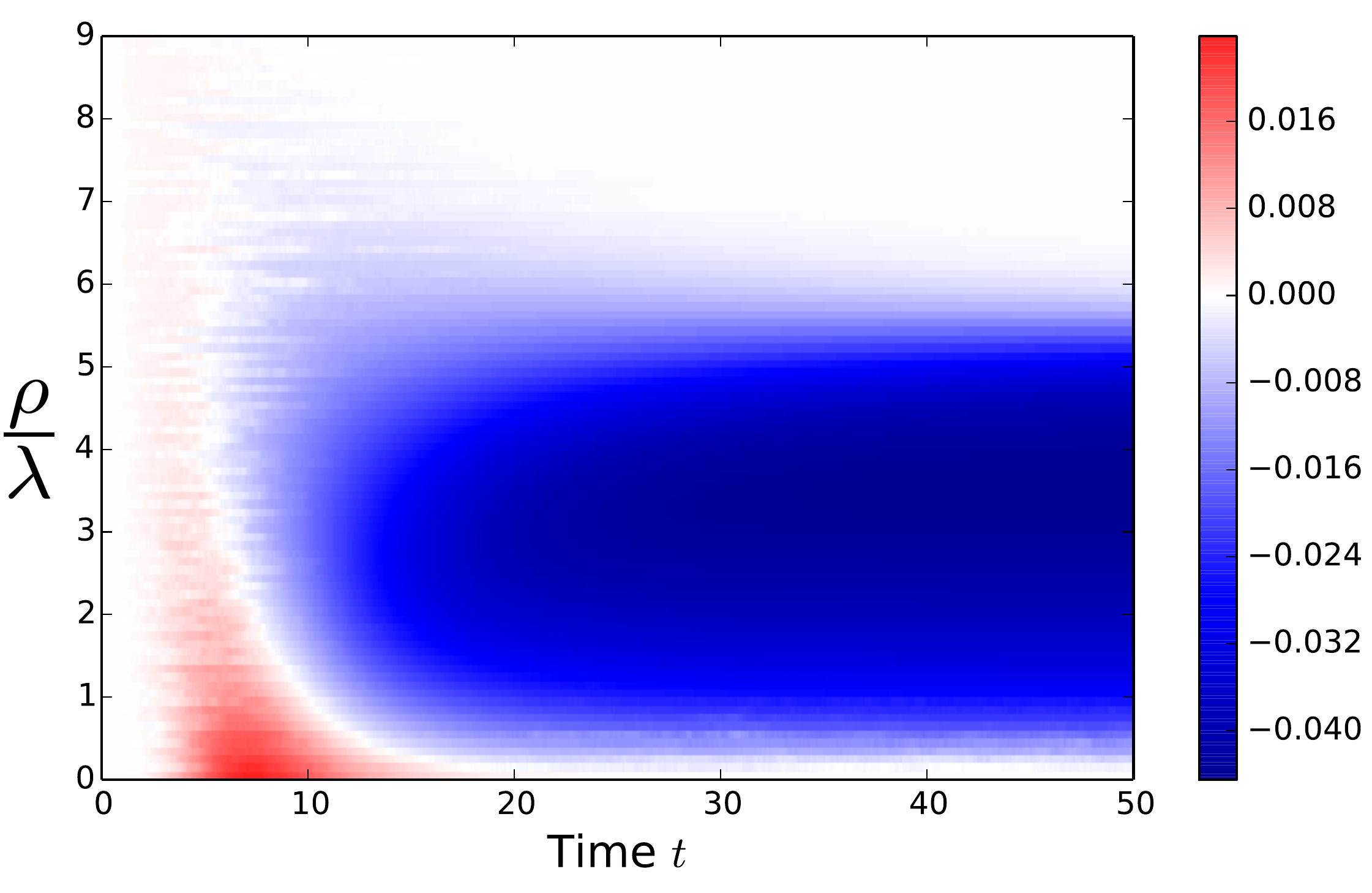}
\caption{The difference between $L_1^{\rDMP}$ and $L_1^\pair$ on Zachary's Karate Club for various values of the ratio $\wane/\lambda$.  We rescale time so that $\lambda = 0.1$ as before.  In the blue region, $L_1^{\rDMP} < L_1^\pair$ and \rDMP\ is more accurate; in the red region, $L_1^{\rDMP} > L_1^\pair$.  We see that \rDMP\ is more accurate except at early times or when $\wane/\lambda$ is small.}
\label{fig:parameter-sweep}
\end{figure}

To investigate how \rDMP\ compares with the pair approximation across a broader range of parameters, in Fig.~\ref{fig:parameter-sweep} we vary the ratio between waning rate $\wane$ and the transmission rate $\lambda$.  Since we can always rescale time by multiplying $\lambda$ and $\wane$ by the same constant, we do this by holding $\lambda = 0.1$ as before, and varying $\wane$.  We then measure the difference in the $L_1$ error of the two methods, $L_1^{\rDMP} - L_1^{\pair}$.  

In the blue region, rDMP is more accurate than the pair approximation; in the red region, it is less so.  We see that \rDMP\ is more accurate except at early times (as in Fig.~\ref{fig:L1distance}) or when $\wane$ is small compared to $\lambda$, i.e., if the model is close to the SI model where Infectious nodes rarely become Susceptible again.  

\begin{figure}
\centering
\includegraphics[width=3.2in]{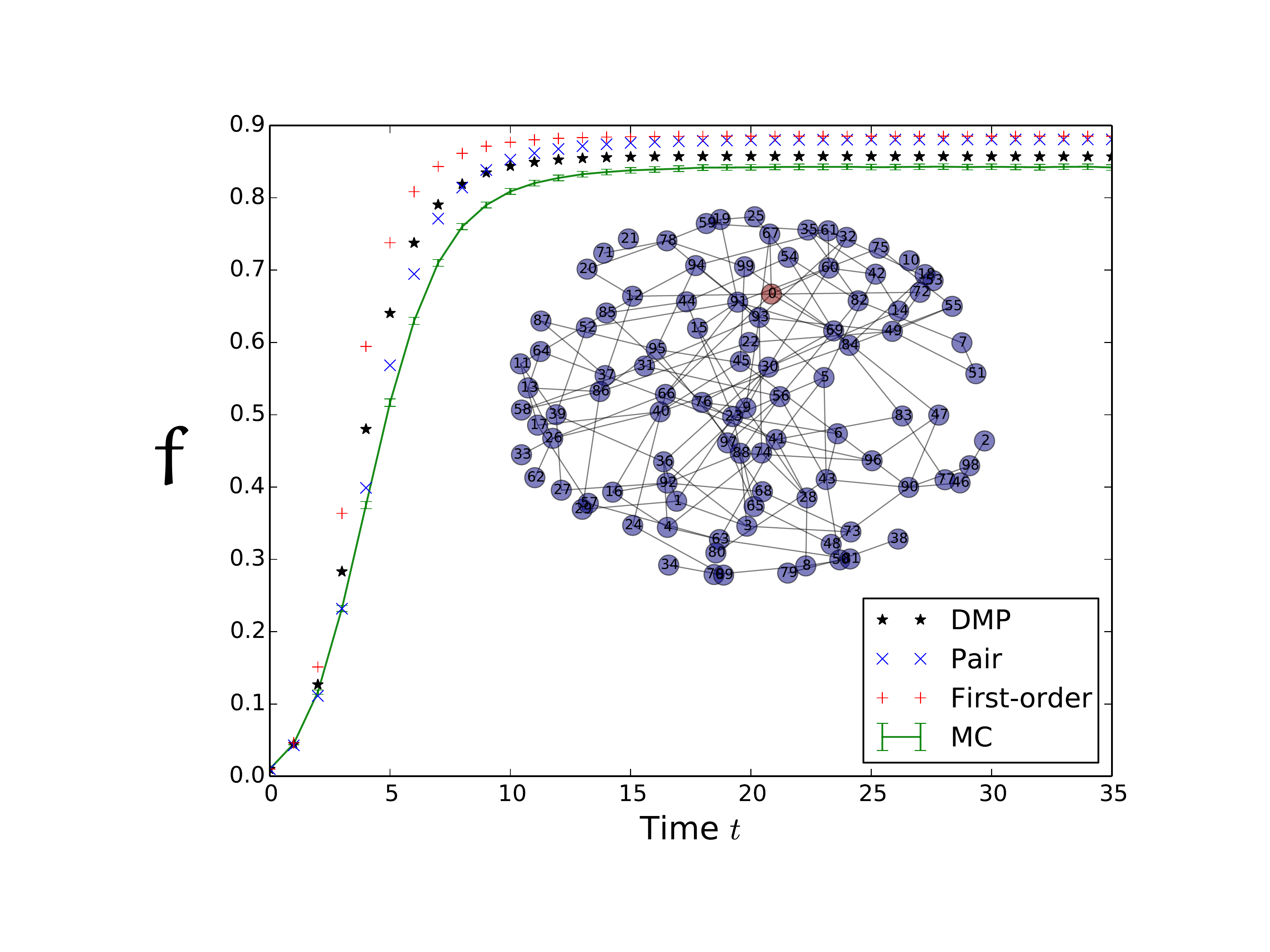}
\caption{The fraction $f$ of Infectious nodes as a function of time in the SIS model on an Erd\H{o}s-R\'enyi graph (inset) with $n=100$ and average degree $3$.  Here $\lambda=0.4$, $\wane = 0.1$, and the initial condition consists of a single Infectious node (colored red).  Monte Carlo results were averaged over $10^3$ independent runs.  Except at early times, \rDMP\ tracks the true trajectory more closely.}
\label{fig:erdos-dynamics}
\end{figure}

In Fig.~\ref{fig:erdos-dynamics}, we simulate the SIS model on an Erd\H{o}s-R\'enyi graph with $n=100$ and average degree $3$, with $\lambda = 0.4$, $\wane = 0.1$, and a single initially Infectious node.  As with the Karate Club, \rDMP\ does a better job of tracking the true fraction of Infectious nodes, except at early times when the pair approximation is superior; in particular, it does a better job of computing the steady-state size of the epidemic.  

\begin{figure*}
\centering
\mbox{\subfigure{\includegraphics[width=3.2in]{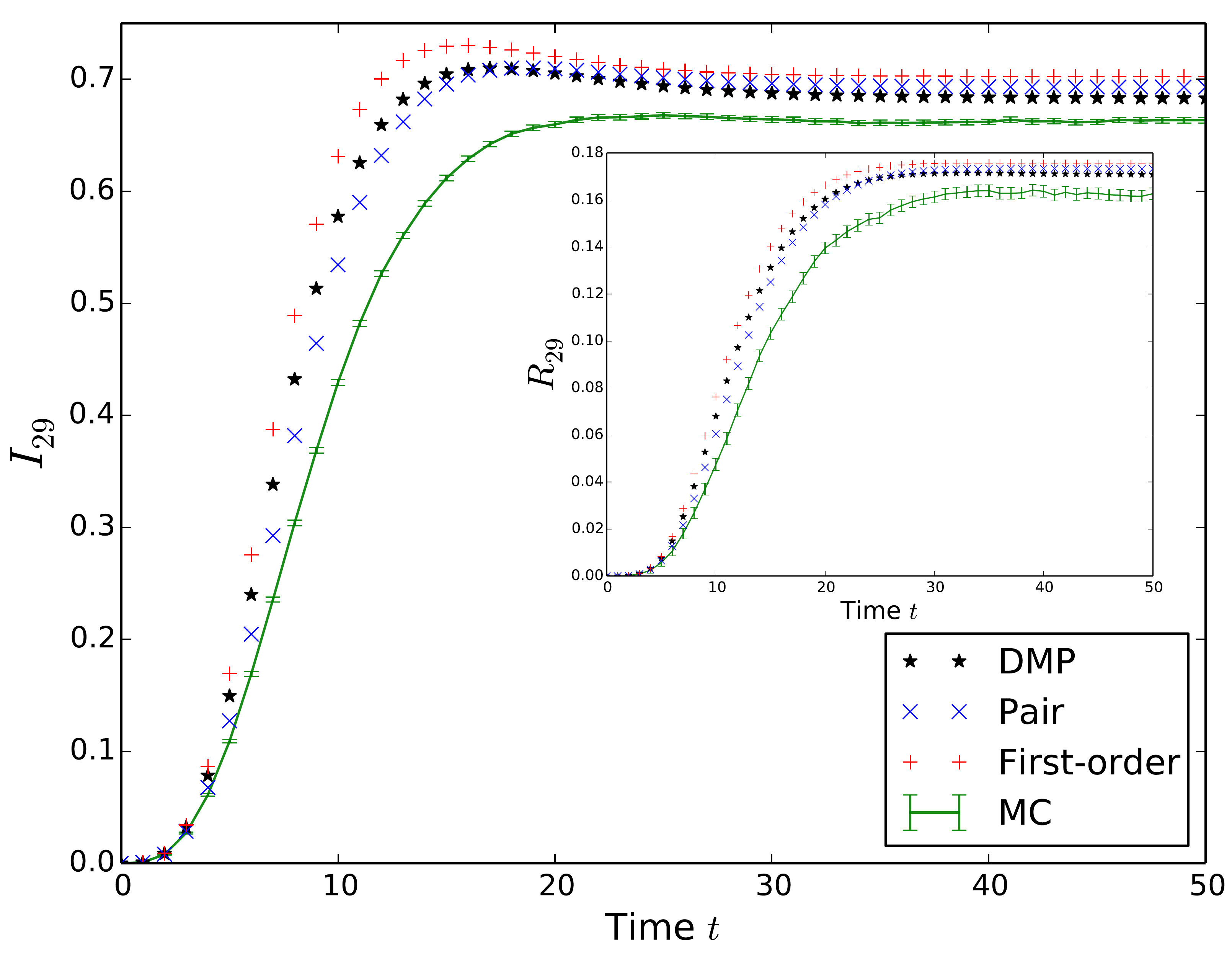}}\quad
\subfigure{\includegraphics[width=3.2in]{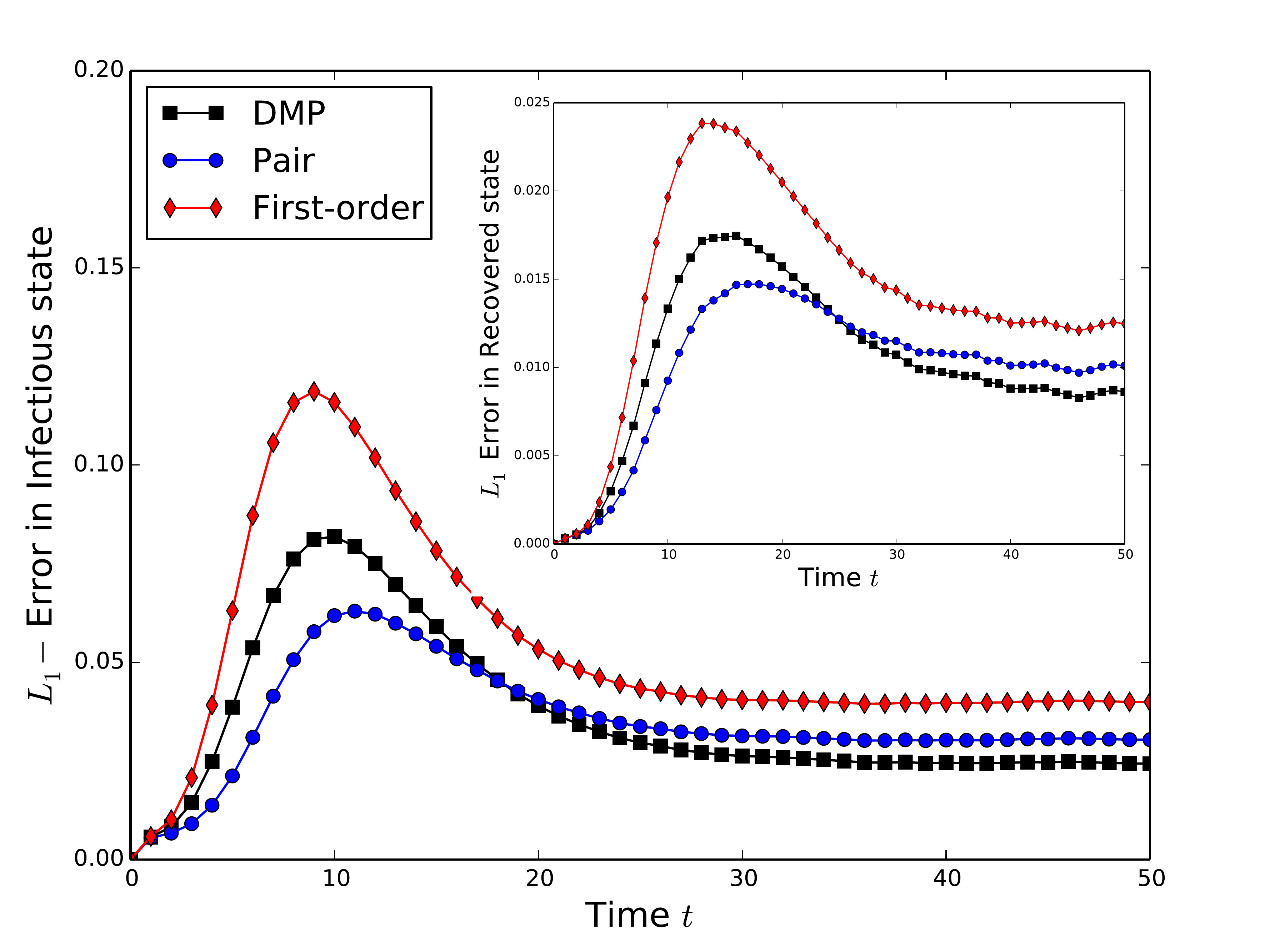}}}
\caption{The SIRS model on the Karate Club.  On the left, we show the true and estimated marginal probability that a node 29 is Infectious (main figure) or Recovered (inset) as a function of time. On the right is the average $L_1$ error for the Infectious and Marginal states.  The transmission rate is $\lambda = 0.1$, and the transition rates from Infectious to Recovered and from Recovered to Susceptible are $\itor = 0.05$ and $\rtos = 0.2$ respectively.  Node $0$ (colored red) was initially infected.  Monte Carlo results were averaged over $10^5$ runs.  As for the SIS model, \rDMP\ is significantly more accurate than the first-order model where nodes are independent, and is more accurate than the pair approximation except at early times.}
\label{fig:error-sirs}
\end{figure*}

In Fig.~\ref{fig:error-sirs} we show results for the SIRS model on Zachary's Karate Club.  As in Fig.~\ref{fig:L1distance}, on the left we show the marginal probability $\infectious_{29}$ that node 29 is Infectious; on the right, we show the $L_1$ error for $\infectious_i$ averaged over the network.  In the insets, we show the marginal probability $\recovered_{29}$ for the Recovered state and the corresponding average $L_1$ error.  Here the transmission rate is $\lambda=0.1$, the waning rate from Infectious to Recovered is $\itor = 0.05$, and the rate from Recovered to Susceptible is $\rtos = 0.2$.  The initial condition consisted of a single infected node, and Monte Carlo results were averaged over $10^5$ runs.  As for the SIS model, \rDMP\ is significantly more accurate than the independent node approximation, and is more accurate than the pair approximation except at early times.  

We found similar results on many other families of networks, including random regular graphs, random geometric graphs, scale-free networks, Newman-Watts-Strogatz small world networks, and a social network of dolphins~\cite{Lusseau}.  Namely, \rDMP\ outperforms the first-order approximation where nodes are independent, and outperforms the pair approximation across a wide range of parameters and times.

\section{Linear Stability, Epidemic Thresholds, and Related Work}
\label{sec:related_works}

Systems of differential equations for \rDMP, such as~\eqref{dmp_edge_sis}, do not appear to have a closed analytic form due to their nonlinearities.  On the other hand, we can compute quantities such as epidemic thresholds by linearizing around a stationary point, such as $\{\infectious_{j \to i}^*=0\}$ where the initial outbreak is small.  Given a perturbation $\epsilon_{j \rightarrow i}=\infectious_{j \to i}-\infectious_{j \to i}^*$, the linear stability of the system, i.e., whether or not $\epsilon_{j \rightarrow i}$ diverges in time, is governed by the eigenvalues of the Jacobian matrix $\J$ of the right hand side of~\eqref{dmp_edge_sis} at the stationary point $ \infectious_i^*$.  The Jacobian for~\eqref{dmp_edge_sis} at $\{\infectious_{j \to i}^*\}$ is 
\be\label{jacobian}
\J_{(j \to i),(k \rightarrow j')} = -\delta_{kj} \delta_{ij'} \rho + \lambda (1- \infectious_{j}^*)  \B_{(j \to i),(k \rightarrow j')} \, . 
\ee
where
\be
 \B_{(j \to i),(k \rightarrow j')}= \delta_{jj'}(1-\delta_{ik}) \, . 
\ee
This definition of $\B$ is another way of saying that the edge $k \to j$ influences edges $j \to i$ for $i \ne k$, but does not backtrack to $k$.  This corresponds to our assumption that infections, for instance, do not bounce from $k$ to $j$ and back again and create an echo chamber effect.  For this reason, $\B$ is also known in the literature as the non-backtracking matrix \cite{Krzakala_Bmatrix} or the Hashimoto matrix \cite{Hashimoto}. 

Now, for a small perturbation $\vec{\epsilon}$ away from a stationary point $\{\infectious_{j \to i}^*\}$, the linearized system of~\eqref{dmp_edge_sis} becomes
\be
 \frac{d\vec{\epsilon}}{dt}= \J \vec{\epsilon}, 
 \ee
If $\J$ has any eigenvalues with positive real part, then $\| \vec{\epsilon}(t) \|$ grows exponentially in time. So, the fixed point $\{\infectious_{j \to i}\}$ is stable as long as the leading eigenvalue $J_{1}$ of $\J$ has negative real part.
 
One trivial, but important, stationary point to test is $\infectious_{j \to i}^*=0$ for all edges.  A small perturbation around $\vec{0}$ corresponds to a small initial probability that each node is infected.  
From~\eqref{jacobian}, $\J$ becomes
\be
\J = \lambda \left( \B-\frac{\wane}{\lambda} \id \right) \, , 
\ee
where $\id$ is the $2m \times 2m$ identity matrix.  So, the leading eigenvalue of $\J$ becomes positive when the largest eigenvalue $B_1$ of $\B$ is greater than $\wane/\lambda$. In other words, if 
\be
\label{eq:b-threshold}
R_0 = \frac{\lambda}{\wane} B_1 \ge 1 \, , 
\ee
where $R_0$ is the reproductive number, even a small initial probability of infection will lead to a widespread endemic state, where the infection becomes extensive.  If~\eqref{eq:b-threshold} does not hold, a small initial probability of infection will instead decay back to an infection-less state.

\begin{figure*}
\centering
\mbox{\subfigure{\includegraphics[width=3.2in]{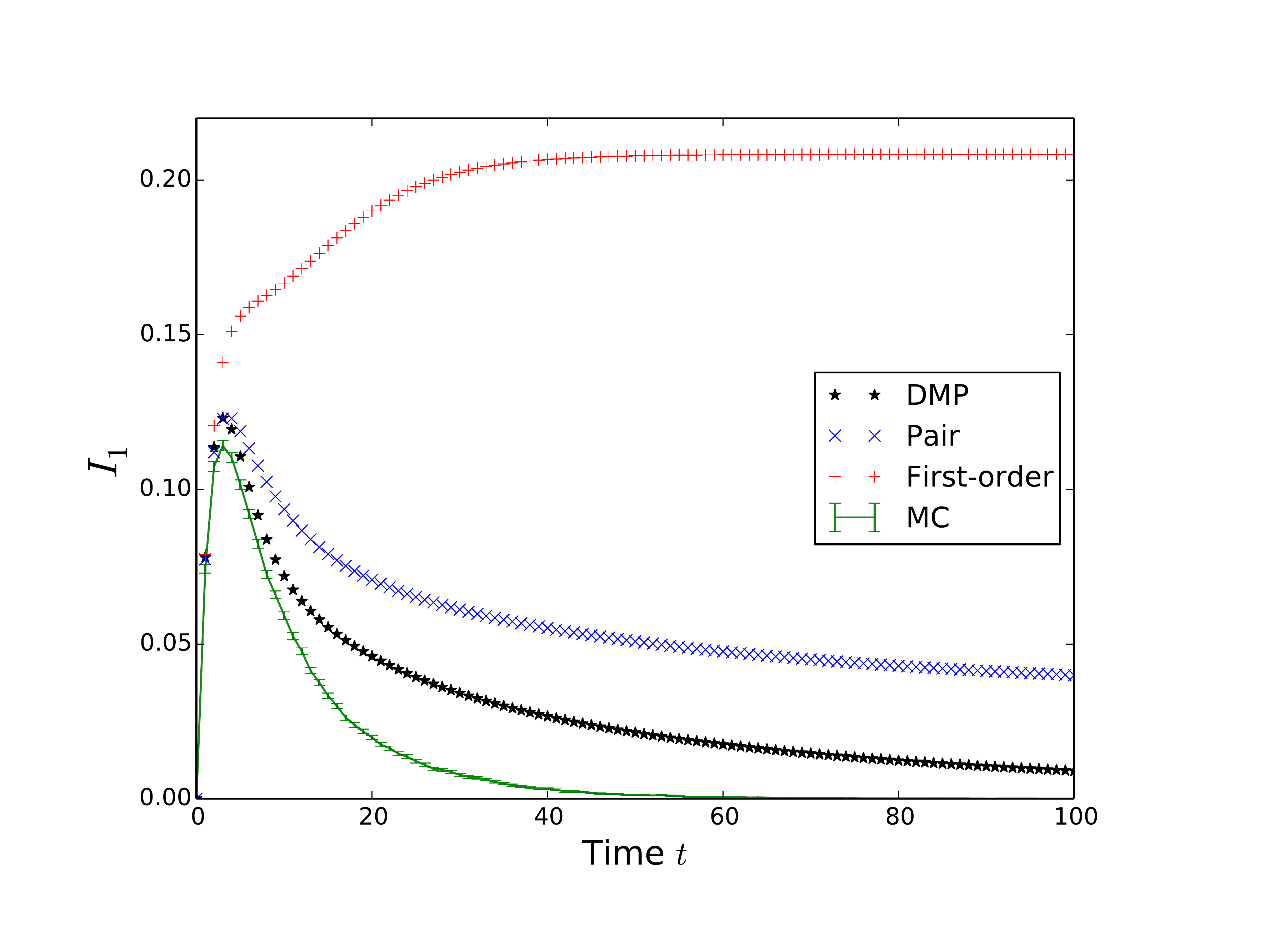}}\quad
\subfigure{\includegraphics[width=3.2in]{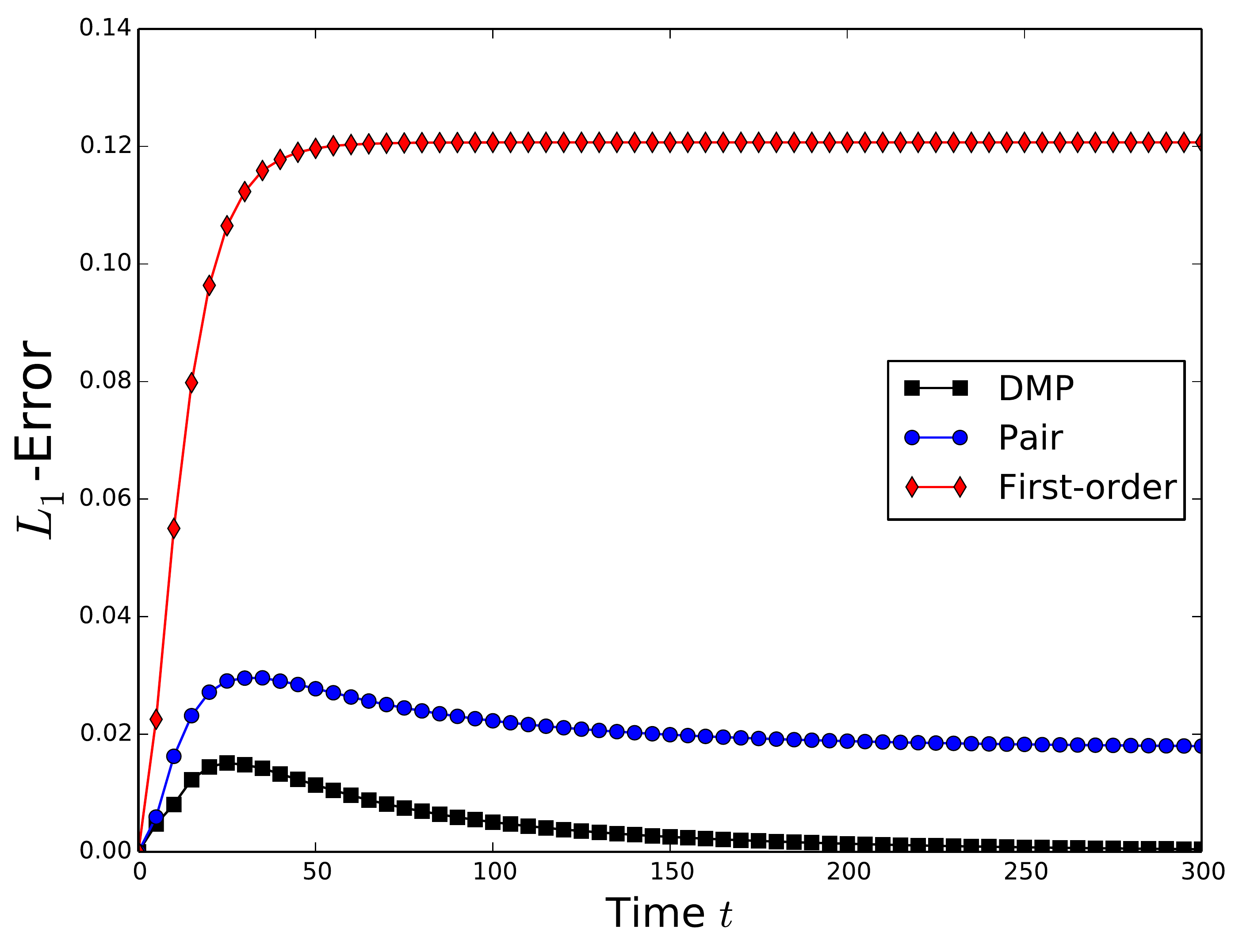} }}
\caption{Same as in Fig.~\ref{fig:L1distance}, but with transmission rate $\lambda = 0.1$ and waning rate $\wane = 0.54$. A well known upper bound on the epidemic threshold of the SIS model can be computed from the leading eigenvalue $A_1$ of the adjacency matrix (the Jacobian matrix of first-moment-closure approach) of a network. In other words, if $\frac{\wane}{\lambda} <A_1$, it is known from the first-moment-method that an infection-free state becomes unstable and epidemics become widespread and endemic. Here we show the results from SIS model in Zachary's Karate Club, where $A_1 \approx 6.7$. Even though  $\frac{\wane}{\lambda} =5.4 <A_1$ which is well below the threshold from the first-moment method, the contagion fades away eventually, which is correctly captured by our DMP approach.}
\label{Node_dynamics_belowcriticality}
\end{figure*}

Since $\B$ is not symmetric, not all its eigenvalues are real.  However, by the Perron-Frobenius theorem, it's leading eigenvalue is real; moreover, it is upper bounded by $A_1$, the leading eigenvalue of the adjacency matrix $\A$.  Interestingly, if we examine the linear stability of the first-order approximation where nodes are independent, 
~\cite{Newman_Network_Book}, the epidemic threshold for the SIS model is given by 
\be
\label{eq:a-threshold}
\frac{\lambda}{\wane} A_1 \ge 1 \, .
\ee
Since $B_1 \le A_1$, the threshold~\eqref{eq:b-threshold} gives a better upper bound for the true epidemic threshold than we would get from the first-order approximation.  
A similar threshold for the SIR model in sparse networks, or equivalently for percolation, using $B_1$ was recently demonstrated in~\cite{Karrer2014}.  (We note that when backtracking is allowed, it has important consequences for epidemic thresholds on power-law networks~\cite{chatterjee2009contact}.) 
 
Whereas the leading eigenvector of $\B$ governs the epidemic threshold, the spectral gap between $\B$'s top two eigenvectors governs how quickly the epidemic converges to the leading behavior (at least until we leave the linear regime).  Qualitatively, this depends on bottlenecks in the network such as those due to community structure, where an epidemic spreads quickly in one community but then takes a longer time to cross over into another.  Indeed, the second eigenvector of the non-backtracking matrix $\B$ was recently used to detect community structure~\cite{Krzakala_Bmatrix}. 

Similarly, just as the leading eigenvector of $\B$ was recently shown to be a good measure of importance or ``centrality'' of a node~\cite{Newman2014}, it may be helpful in identifying ``superspreaders''---nodes where an initial infection will generate the largest outbreak, and be the most likely to lead to a widespread epidemic.

\section{Conclusion}
\label{sec:conclusion}

Modern epidemiological studies often require recurrent models, where nodes can return to their previous inhabited states multiple times.  For example, consider diseases such as influenza where 
individuals are infected multiple times throughout their lives, or whooping cough where vaccine effectiveness wanes over time; in both cases, individuals return to the Susceptible class.  In this paper we have extended Dynamic Message-Passing (DMP) to recurrent epidemic models.  Our \rDMP\ approach defines messages on the directed edges of a network in such a way as to prevent signals, such as the spread of infection, from backtracking immediately to the node that they came from.  By preventing these ``echo chamber effects,'' \rDMP\ obtains good estimates of the time-varying marginal probabilities on a wide variety of networks, estimating both the fraction of infectious individuals in the entire network, and the probabilities that individual nodes become infected.

Like the pair approximation, \rDMP\ takes correlations between neighboring nodes into account.  However, our experiments show that \rDMP\ is more accurate than the pair approximation for a wide variety of network structures and parameters.  Moreover, \rDMP\ is computationally less expensive than the pair approximation, especially for complex epidemic models with a large number of states, using $O(mk)$ instead of $O(mk^2)$ variables for models with $k$ states on networks with $m$ edges.  

Finally, \rDMP\ is conceptually simple, allowing the user to immediately write down the system of differential equations for a wide variety of epidemic models, such as those with multiple stages of infection or immunity \cite{melnik2013multistage,miller2013aa}, or those with multiple interacting diseases \cite{karrer2011competing,miller2013cocirculation}. We expect that given its simplicity and accuracy, it will be an attractive option for future epidemiological studies.

\section{Acknowledgments}
This work is supported by AFOSR and DARPA under grant \#FA9550-12-1-0432.  MS performed this work while a Graduate Fellow at the Santa Fe Institute, and SVS was supported by the Santa Fe Institute and the Omidyar Group.  We are grateful to Mason Porter and Joel Miller for helpful conversations regarding recurrent state epidemic models.


\begin{thebibliography}{99} 

\bibitem{KarrerNewman_2010}
 B. Karrer and M.E.J. Newman,  
\newblock Message passing approach for general epidemic models. 
\newblock \emph{Phys. Rev. E} {\bf 82}, 016101 (2010)

\bibitem{miller_2011}
Joel C. Miller, Anja C. Slim and Erik M. Volz, 
\newblock Edge-based compartmental modelling for infectious disease spread. 
\newblock \emph{Journal of the Royal Society Interface [Internet].} {\bf 9} 890-906 (2010).

\bibitem{ShresthaMoore_2014_1}
 M. Shrestha and C. Moore,  
\newblock Message passing approach for threshold models of behavior in networks.
\newblock \emph{Phys. Rev. E} {\bf 89}, 022805 (2014)

\bibitem{Altarelli_2013_L}
F. Altarelli, A. Braunstein, L. Dall'Asta, and R. Zecchina,  
\newblock Large deviations of cascade processes on graphs.
\newblock \emph{Phys. Rev. E} {\bf 87} 062115 (2013)

\bibitem{Lokhov_2014}
A.Y. Lokhov, M. M\'ezard, and L. Zdeborov\`a, 
\newblock Dynamic message-passing equations for models with unidirectional dynamics.
\newblock \emph{Phys. Rev. E} {\bf 91}, 012811 (2015)

\bibitem{Bailey75}
N.~T.~J. Bailey, \emph{The Mathematical Theory of Infectious Diseases and its
  Applications}. Hafner Press, New York (1975).

\bibitem{AndersonMay91}
R.~M. Anderson and R.~M. May, 
\emph{Infectious Diseases of Humans}. Oxford University Press, Oxford (1991).

\bibitem{Gran78}
M. Granovetter, 
\newblock Threshold models of collective behavior.
  \emph{American Journal of Sociology} \textbf{83(6)}, 1420�1443(1978).
  
\bibitem{Gran73}
\newblock M. Granovetter, 
\newblock The strength of weak ties.
 \emph{American Journal of Sociology} \textbf{78(6)}, 1360�1380(1973).

\bibitem{JMiller04}
J.H. Miller and S.E. Page, 
\newblock The standing ovation problem.
 \emph{Complexity} \textbf{9}, 8-16 (2004).  

\bibitem{Goncalves}
B. Gon\c calves, N. Perra, A. Vespignani, 
\newblock Modeling Users' Activity on Twitter Networks: Validation of Dunbar's Number.
\newblock {\emph{PLoS ONE }  {\bf 6 (8)}}, e22656 (2011).

\bibitem{BMay2011}
R.~M. May and A.~G. Haldane, 
\newblock Systemic risk in banking ecosystems. 
\newblock \emph{Nature}  \textbf{469}, 351-355 (2011).

\bibitem{CSMF2012}
F. Caccioli, M. Shrestha, C. Moore, and J. D Farmer, Stability analysis of financial contagion due to overlapping portfolios.
\newblock {\emph{Journal of Banking \& Finance} {\bf 46}},  233-245 (2014).

\bibitem{BakChen1}
P. Bak, K. Chen, and C. Tang, A forest-fire model and some thoughts on turbulence. 
\newblock {\emph{Phys. Lett. A,} {\bf 147}},  297-300  (1990).

\bibitem{Drossel1}
B. Drossel, and F. Schwabl, Self-organized critical forest-fire model.
\newblock {\emph{Phys. Rev. Lett.} {\bf 69}}, 1629-1632 (1992). 

\bibitem{Grassberger1}
P. Grassberger, 
\newblock Critical behaviour of the Drossel-Schwabl forest fire model. 
\newblock {\emph{New J. Phys,} {\bf 4}},  17 (2002).

\bibitem{MezardMontanari}
M. M\'ezard and A. Montanari, 
\newblock \emph{Information, Physics, and Computation.}
\newblock Oxford University Press (2009).

\bibitem{MooreMertens}
C. Moore and S. Mertens,
\newblock \emph{The Nature of Computation.}
\newblock Oxford University Press (2011).

\bibitem{RMorris1}
R. Morris, Zero-temperature Glauber dynamics on $\mathrm{Z}^d$. 
\newblock {\emph{Prob. Theory Rel. Fields,} {\bf 149}}, 3-4   (2011).

%
%
%
%


\bibitem{Dunbar}
R.I.M Dunbar, 
\newblock Neocortex size as a constraint on group size in primates.
\newblock \emph{Journal of Human Evolution}  {\bf 22 (6)}, 469-493 (1992)

\bibitem{Meyers2007}
L. A. Meyers,
\newblock Contact network epidemiology: Bond percolation applied to infectious disease prediction and control,
\newblock \emph{Bulletin of the American Mathematical Society} {\bf 44} 63-86 (2007).

\bibitem{Newman_Network_Book}
M. E. J. Newman,
\newblock \emph{Networks: An Introduction}.
\newblock {Oxford University Press} (2010).



\bibitem{Lokhov_2013}
A.Y. Lokhov, M. M\'ezard, H. Ohta, and L. Zdeborov\`a,  
\newblock Inferring the origin of an epidemic with dynamic message-passing algorithm.
\newblock \emph{Phys. Rev. E} {\bf 90}, 012801 (2014)

\bibitem{Altarelli_2014_1}
F. Altarelli, A. Braunstein, L. Dall'Asta, A. Ingrosso, and R. Zecchina,  
\newblock The zero-patient problem with noisy observations.
\newblock \emph{J. Stat. Mech} P10016 (2014)

\bibitem{Altarelli_2013_1}
F. Altarelli, A. Braunstein, L. Dall'Asta, J.R. Wakeling, and R. Zecchina,  
\newblock Containing epidemic outbreaks by message-passing techniques.
\newblock \emph{Phys. Rev. X} {\bf 4} 021024 (2014)

\bibitem{Altarelli_2013_2}
F. Altarelli, A. Braunstein, L. Dall'Asta, and R. Zecchina,  
\newblock Optimizing spread dynamics on graphs by message passing.
\newblock \emph{J. Stat. Mech} P09011 (2013)

\bibitem{Altarelli_2011_1}
F. Altarelli, A. Braunstein, A. Ramezanpour, and R. Zecchina,  
\newblock Stochastic optimization by message passing.
\newblock \emph{J. Stat. Mech}  P11009 (2011)

\bibitem{Mason-review-2015}
 M. A. Porter and J. P. Gleeson,
\newblock Dynamical systems on networks: A tutorial.
\newblock {arXiv:1403.7663} (2014).


\bibitem{Lusseau}
D. Lusseau, K. Schneider, O. J. Boisseau, P. Haase, E. Slooten, and S. M. Dawson,
\newblock The bottlenose dolphin community of Doubtful Sound features a large proportion of long-lasting associations.
\newblock \emph{Behavioral Ecology and Sociobiology} {\bf 54}, 396-405 (2003).

\bibitem{ZhangMoore_2014_1}
P. Zhang, and C. Moore,
\newblock Scalable detection of statistically significant communities and hierarchies: message-passing for modularity.
\newblock \emph{Proceedings of the National Academy of Sciences } {\bf 111 } (51), 18144-18149


\bibitem{Hashimoto}
K. Hashimoto,
\newblock Zeta functions of finite graphs and representations of $p$-adic groups. 
\newblock  \emph{Advanced Studies in Pure Mathematics }  {\bf 15} 211-280 (1989).

\bibitem{JPearl1}
J. Pearl, 
\newblock  Reverend Bayes on inference engines: a distributed hierarchical approach.
 \emph{AAAI Proceedings} \textbf{82}, (1982).

  \bibitem{Decelle2011}
A. Decelle, F. Krzakala, C. Moore, and L. Zdeborov\'a, Asymptotic analysis of the stochastic block model for modular networks and its algorithmic applications.
\newblock {\emph{Phys. Rev. E} {\bf 84}}, 066106 (2011).

\bibitem{Zachary}
W. W. Zachary,
\newblock An information flow model for conflict and fission in small groups. 
\newblock \emph{Journal of Anthropological Research} {\bf 33 (4)}, 452-473 (1977).
  
\bibitem{KeelingRohani08}
 M. J. Keeling and P. Rohani, \emph{ Modeling Infectious Diseases in Humans and Animals}. Princeton and Oxford: Princeton University Press (2008).

\bibitem{Krzakala_Bmatrix}
F. Krzakala, C. Moore, E. Mossel, J. Neeman, A. Sly, L. Zdeborov\'a, and P.  Zhang,\
\newblock Spectral redemption in clustering sparse networks. 
\newblock \emph{Proceedings of the National Academy of Sciences} {\bf 110 (52)}, 20935-20940 (2013).

\bibitem{Karrer2014}
B. Karrer, M. E. J. Newman, and L. Zdeborov\'a,
\newblock Percolation on sparse networks. 
\newblock \emph{Phys. Rev. E} {\bf 113}, 208702 (2014).

\bibitem{chatterjee2009contact}
S. Chatterjee and R. Durrett,
\newblock Contact processes on random graphs with power law degree distributions have critical value 0. 
\newblock \emph{The Annals of Probability} {\bf 37}, 2332--2356 (2009).


\bibitem{Watson_galton}
H W Watson, and Francis Galton, 
\newblock On the Probability of the Extinction of Families
\newblock\emph{Journal of the Anthropological Institute of Great Britain},  {\bf 4},  138-144, (1875).

\bibitem{Newman2014}
T. Martin, X. Zhang, M. E. J. Newman,
\newblock Localization and centrality in networks. 
\newblock \emph{Phys. Rev. E} {\bf 90}, 052808 (2014).

\bibitem{earn2002ecology}
D. J. D Earn, J Dushoff, S. A Levin,
\newblock Ecology and evolution of the flu.
\newblock \emph{Trends in ecology \& evolution} {\bf 17}, 334--340 (2002).

\bibitem{gomes2004infection}
M. G. M Gomes, L. J White, G. F Medley,
\newblock Infection, reinfection, and vaccination under suboptimal immune protection: epidemiological perspectives.
\newblock \emph{Journal of Theoretical Biology} {\bf 228}, 539--549 (2004).

\bibitem{melnik2013multistage}
S. Melnik, J. A. Ward, J. P. Gleeson, and M. A. Porter, 
\newblock Multi-stage complex contagions. 
\newblock \emph{Chaos} {\bf 23}, 013124 (2013).

\bibitem{miller2013aa}
J. C. Miller and E. M Volz,
\newblock Incorporating Disease and Population Structure into Models of SIR Disease in Contact Networks. 
\newblock \emph{PLoS ONE} {\bf 8}, (8) e69162 (2013).

\bibitem{karrer2011competing}
B. Karrer and M. E. J. Newman,
\newblock Competing epidemics on complex networks,
\newblock \emph{Phys. Rev. E} {\bf 84}, 036106 (2011).

\bibitem{miller2013cocirculation}
J. C. Miller,
\newblock Cocirculation of infectious diseases on networks, 
\newblock \emph{Phys. Rev. E} {\bf 87}, 060801 (2013).

\bibitem{golden2005effect}
M. R. Golden, W. L. H. Whittington, H. H. Handsfield, J. P. Hughes, W. E. Stamm, M. Hogben, A. Clark, C. Malinski,  J. R. L Helmers,  K. K. Thomas,  and K. K Holmes,
\newblock Effect of expedited treatment of sex partners on recurrent or persistent gonorrhea or chlamydial infection. 
\newblock \emph{New England Journal of Medicine} {\bf 352}, 676--685 (2005).

\bibitem{conway2007recurrent}
P. H. Conway, A. Cnaan, T. Zaoutis, and B. V. Henry, R. W. Grundmeier, and R. Keren,
\newblock Recurrent urinary tract infections in children: risk factors and association with prophylactic antimicrobials. 
\newblock \emph{Journal of the American Medical Association} {\bf 2}, 179--186 (2007).

\bibitem{jeffery1966epidemiological}
G. M. Jeffery,
\newblock Epidemiological significance of repeated infections with homologous and heterologous strains and species of Plasmodium. 
\newblock \emph{JBulletin of the World Health Organization} {\bf 35}, 873 (1966).

\bibitem{drusin2000nosocomial}
L. M. Drusin, B. G. Ross, K. H. Rhodes, A. N. Krauss, R. A. Scott, 
\newblock Nosocomial Ringworm in a Neonatal Intensive Care Unit A Nurse and Her Cat. 
\newblock \emph{Infection Control} {\bf 21}, 605--607 (2000).



\end{thebibliography}
\end{document}